\documentclass[superscriptaddress,twocolumn ]{revtex4}
\usepackage{epsfig,amsmath,graphicx,amssymb}
\usepackage[colorlinks=fals,linkcolor=blue]{hyperref}

\def\be{\begin{equation}}
\def\ee{\end{equation}}
\def\bea{\begin{eqnarray}}
\def\eea{\end{eqnarray}}

\begin{document}

\title{{\large Manipulate the coiling and uncoiling movements of Lepidoptera proboscis
by its conformation optimizing} }
\author{Xiaohua Zhou}
\email{xhzhou08@gmail.com}
 \affiliation{Department of Applied
Physics, Xi'an Jiaotong University, Xi'an 710049, People's Republic
of China}
 \affiliation{Department of
Mathematics and Physics, Fourth Military Medical University, Xi'an
710032, People's Republic of China }
\author{Shengli Zhang}
\email{zhangsl@mail.xjtu.edu.cn}
\affiliation{Department of Applied Physics, Xi'an Jiaotong University, Xi'an 710049,
People's Republic of China}
\date{\today}

\begin{abstract}
Many kinds of adult Lepidoptera insects possess a long proboscis
which is used to suck liquids and has the coiling and uncoiling
movements. Although experiments revealed qualitatively that the
coiling movement is governed by the hydraulic mechanism and the
uncoiling movement is due to the musculature and the elasticity, it
needs a quantitative investigation to reveal how insects achieve
these behaviors accurately. Here a quasi-one-dimensional (Q1D)
curvature elastica model is proposed to reveal the mechanism of
these behaviors. We find that the functions of internal stipes
muscle and basal galeal muscle which locate at the bottom of
proboscis are to adjust the initial states in the coiling and
uncoiling processes, respectively. The function of internal galeal
muscle which exists along proboscis is to adjust the line tension.
The knee bend shape is due to the local maximal spontaneous
curvature and is an advantage for nectar-feeding butterfly. When
there is no knee bend, the proboscis of fruit-piercing butterfly is
easy to achieve the piercing movement which induced by the increase
of internal hydraulic pressure. All of the results are in good
agreement with experiential observation. Our study provides a
revelatory method to investigate the mechanical behaviors of other
1D biologic structures, such as proboscis of marine snail and
elephant. Our method and results are also significant in designing
the bionic devices.

\textbf{keywords:} {Curvature, Proboscis, Equilibrium shape}
\end{abstract}

\pacs{XX, XX}
\maketitle


\section{ Introduction}
 After millions of years of evolution, many lives
possess particular biological functions to adapt the environment.
These adaptive capacities usually present as predominance to capture
food, defend enemy and disease, obtain isomerism, and so on. For
instances, the lotus leaves \cite{Neinhuis,Lafuma} have the
self-cleaning ability \textrm{due to} \textrm{plentiful} micro- and
nano-protrusions on the surface and similar structure also be found
on the feet of geckoes \cite{Hansen}. The hummingbird uses the
liquid-trapping method that changes configuration and shape of the
tongue dramatically to pick up liquids because it is highly
efficient \cite{Alejandro}. The above phenomena indicate that the
morphogenesis of structures in biological system govern biological
functions. So research on the conformations of biological organs can
give a deep insight to understand the relationships between their
geometric shapes and biological functions, and it also can give us
inspiration in designing the bionic devices.
\begin{figure}\centering
\includegraphics[width=0.45\textwidth]{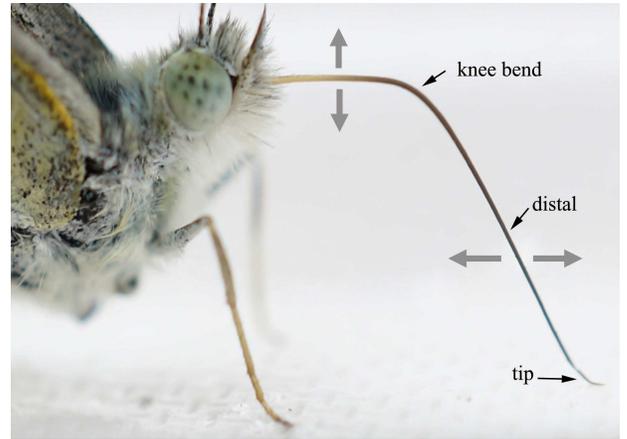} \caption{\label{fig1} The
proboscis of a nectar-feeding butterfly (Pieris canidia) has the
knee bend and has two particular movements: the up-and-down movement
of whole proboscis and the front-and-back movement of the distal
region.}
\end{figure}

The Lepidoptera (butterflies and moths) are one of the most diverse
taxa of
animals with about 160,000 described species contained in 47 superfamilies %
\cite{Kristensen2007}. The adult Lepidoptera often have a long
proboscis
that is used to suck floral nectar and other liquid substances \cite%
{Krenn2008}. The proboscis usually is coiled like a clockwork
spring. When it needs to suck, the coiled proboscis can uncoil and
extend to a
considerable length, and after that it uncoils to the initial condition \cite%
{Krenn1990}. The coiling and uncoiling movements of proboscis
attract many researchers' attention
\cite{Kristensen1968,Banziger,Krenn2000,Krenn2010}. Experiments
observed that, although the proboscises of different species have
different length and muscle distribution, they have the same
mechanism movements: hydraulic pressure and muscle tensile force
drive the coiling and uncoiling movements
\cite{Krenn2000,Krenn2010}. However, proboscises of different
species also have difference. See Fig.~\ref{fig1}, the extended
proboscis of nectar-feeding butterfly has a knee bend and has two
particular movements: the up-and-down movement of whole proboscis
and the front-and-back movement of the distal region
\cite{Krenn2008,Krenn2010}. The proboscis of fruit-piercing
butterfly has no evident knee bend, but it has another ability:
piercing \cite{Krenn2008,Krenn2010}. What principle dominates these
phenomena and how the butterflies to manipulate their proboscis are
significant and challenging to understand the mechanism of proboscis
movement quantitatively.

In this paper, we construct a quasi-one-dimensional (Q1D) curvature
elastic model to study the movements of proboscis theoretically.
Making use of the energy-minimizing principle, we derive an
equilibrium shape equation of the proboscis. Four typical movements
in one circulation of the coiling and uncoiling movements of
proboscis are achieved, which are in good agreement with
experimental observations. Our results reveal that the knee bend
which is achieved by a local maximal spontaneous curvature is a
disadvantage if the proboscis is needed to pierce into a target. Our
findings not only provide an insight into the mechanics behaviors of
proboscis, but also can give us elicitation to study other
biological behaviors, such as the movement of proboscis of marine
snail and elephant, and the growth of twigs of plants and hairs.

\section{ Model}

The long proboscis often has a much bigger length wide ratio and
although its cross section area is variational along its central
line \cite{Krenn1990,Krenn2000,Krenn2010}, there is no
evident difference of microstructure on different cross section \cite%
{Krenn2000}, which make it can be taken as isotropic 1D material.
However, the volume and corresponding volumetric energy of proboscis
can not be ignored due to there is hydraulic pressure in it
\cite{Krenn2000}. So, we look it as a Q1D elastic structure. An
uncoiling process needs about 1 second and a coiling process needs
about 2.5 seconds \cite{Krenn1990}. We suppose one circulation is a
long time so that each state in whole process can be taken as an
equilibrium configuration.

Considering most of movements of proboscis are in a plane, we
simplify its motion in the planar case. Taking the central line of
proboscis as a 1D elastic structure in the $x-y$ plane, letting $Y$
be the Young's modulus and $I_{z}$ be the moments of inertia of the
cross section around the $z$ axis, we model the elastic energy
density for proboscis
\bea\label{model}
\mathcal {F}=\frac{1}{2}\kappa (K-C)^2+f-\Delta P \sigma_a,
\eea
where $\kappa =YI_{z}$ is the bending rigidity, $C$ is the
spontaneous curvature and $f$ is the inner line tension which
depends on the contraction of muscles along proboscis, $\Delta
P=P_{in}-P_{out}$ is the pressure difference between the inner and
outer of proboscis and $\sigma_a$ is the empty area of the cross
section of proboscis. On the right hand side of the Eq.~\ref{model},
the first and second terms compose the typical 1D model \cite{Tu,
Shi}, but the last term which derivers from the volumetric energy
density makes it should be taken as Q1D model. Traditionally,
$\kappa $, $C$, $f$ and $\sigma_a$ are taken as invariable
constants. But if the cross section area of a Q1D structure is
variational along the arc length of its central line: $s$, such as
proboscis has a decrescent cross section area departing from its
bottom to tip \cite{Krenn2000}, we should choose $\kappa =\kappa
(s)$. Here we choose a more general case that $\kappa =\kappa (s)$,
$C=C(s)$, $f=f(s)$ and $\sigma_a =\sigma_a(s)$ depend on $s$.

The general equilibrium shape equations for the model in
Eq.~\ref{model} have been studied by many researchers
\cite{Capovilla,Thamwattana} and similar models were used to
investigate DNA, filaments and climbing plants
\cite{Benham,Goldstein, Kessler,Bojan,Goriely}. Letting
$\psi=\psi(s)$ be the angle between the $x$
axis and the tangent to the central line of proboscis, we have $K=d\psi /ds=%
\dot{\psi}$ and the equilibrium shape equation for the model in
Eq.~\ref{model} becomes
\bea\label{eq}
\nonumber
2\ddot{\kappa}(\dot{\psi}-C)+4\dot{\kappa}(\ddot{\psi}-\dot{C})+2\kappa(\dddot{\psi}-\ddot{C})
+\kappa \dot{\psi}(\dot{\psi}^2-C^2)\\
-2(f+\Delta P \sigma_a) \dot{\psi}=0.
\eea
The expression of $\kappa =\kappa (s)$, $C=C(s)$, $f=f(s)$ and
$\sigma_a =\sigma_a(s)$ depend on the structure of proboscis.

\begin{figure}\centering
\includegraphics[width=0.45\textwidth]{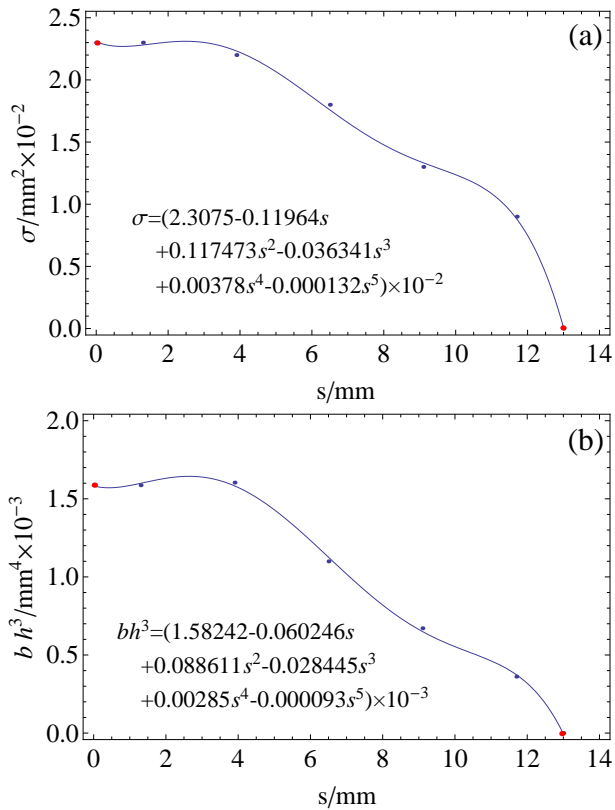}
\caption{\label{fig2} The fitting curves of the configuration
parameters of proboscis with the total length $s=13$ mm. The point
values come form table 1 of ref.~\cite{Krenn2000}. (a) The
relationship of the cross section area $\sigma$ and the arc length
$s$. (b) The fitting curve of $bh^3$ and the arc length $s$. The two
initial points in red color at $s=0$ are used the values of the two
second points at $s=1.3$ mm, respectively. At the two end points we
choose $\sigma=0$ and $bh^3=0$.}
\end{figure}

In ref.~\cite{Krenn2000}, the author measured 24 proboscises came
form the same kind of butterfly, and obtained that the average total
length is about $13$ mm and also provided the cross section area at
different length proportions. We simulated the results and obtained
the cross section area $\sigma=\sigma(s)$ shown in
Fig.~\ref{fig2}(a). Here we should note the empty area $\sigma_a$
should be smaller than $\sigma$ at the same $s$. Choosing
$\sigma_a=\epsilon\sigma$ where the constant $0<\epsilon<1$ and
considering there is always the multiplier $\Delta P$ before
$\sigma_a$ in Eq.~\ref{eq}, the constant $\epsilon$ can be absorbed
in $\Delta P$ and we can simply choose $\sigma_a=\sigma$.

For simplicity, we choose $Y=1$ and have $\kappa =I_{z}$. The
inertia moment $I_{z}$ which depends on the irregular cross section
of proboscis \cite{Krenn2000}. Considering a 1D structure which has
a rectangle cross section with the high $h$ and width $b$, when
bending it
around its central line which parallels to the width, the inertia moment is $%
I=\frac{1}{12}bh^{3}$. If the cross section is a ring with the
radius $R$, we have $I=\frac{\pi }{4}R^{4}=\frac{\pi }{4}RR^{3}$.
Comparing the above two results, if the cross section is irregular,
it is reasonable that the inertia moment can be written as
$I=\lambda bh^{3}$, where $h$ is high, $b$ is width and $\lambda $
is a shape factor which is determined by the shape of cross section.
Experimental results indicate that the cross
section of proboscis at different length proportions are similar \cite%
{Krenn2000}, so we can think the shape factor $\lambda $ is a
constant and have $I_{z}=\lambda bh^{3}$. Making use of the
experimental data in ref.~\cite{Krenn2000}, we obtained the
relationship between $bh^{3}$ and $s$ shown in Fig.~\ref{fig2}(b).
In our calculation we choose $\lambda =1$ and $I_{z}=bh^{3}$.
Clearly, $Y=1$ and $\lambda =1$ are corresponding to the parameters
change: $ F\rightarrow \frac{\mathcal{F}}{Y\lambda }$, $\kappa
\rightarrow \frac{\kappa }{Y\lambda }$, $f\rightarrow
\frac{f}{Y\lambda }$
 and $\Delta P\rightarrow \frac{\Delta P}{Y\lambda }$. In the
following text we will use these changed parameters.

It is found that the butterfly proboscis has a loosely coiled state
which only depends on the elasticity of the proboscis
\cite{Banziger,Krenn1990,Krenn2000,Krenn2005}. For the
nectar-feeding \textrm{butterfly}, there are an evident knee bend
when proboscis uncoiling \cite{Krenn2000,Krenn2008,Krenn2010}.
Experiment results indicate there isn't evident deference at the
knee bend region \cite{Krenn2000}. In our model we think these
phenomena are due to the spontaneous curvature $C$. However, to
precisely obtain it by measurement is difficult. Here we
\textrm{choose} a simpler form
\bea
\label{SC1} C=C_2\exp[-(s-s_0)^2/\tau_1]+C_1\exp[s/\tau_2]+C_0.
\eea
where $C_{0}$, $C_{1}$, $C_{2}$, $\tau _{1}$, $\tau _{2}$ and
$s_{0}$ are constants which are determined by the microcosmic
structure of proboscis. In the above equation, the first term on the
right hand side has distribution feature and it has a local maximum
at $s=s_{0}$. We will see that this term will induce the knee bend
shape when the proboscis uncoiling. The second and third terms on
the left hand side will make sure the proboscis is coiled
with few turns in the loosely coiled state. In our calculation we choose $%
C_{0}=0.55$ mm$^{-1}$, $C_{1}=0.05$ mm$^{-1}$, $C_{2}=0.3$
mm$^{-1}$, $\tau _{1}=1$ mm$^{2}$, $\tau _{2}=4$ mm and $s_{0}=4.44$
mm which is $35\%$ of the proboscis length \cite{Krenn2000}.

Along proboscis it distributes many internal galeal muscles and
their function is to coil proboscis from the loosely coiled state
with few turns to the tightly coiled resting position with more
turns \cite{Krenn2000}. Considering the internal galeal muscle has
degressive cross section area, we choose the following gradually
decreasing line tension which can be own to the constriction of
internal galeal muscles
\bea
\label{f}      f=f_1\exp[-s/\tau_3],
\eea
where $f_{1}$ is a constant which can be adjusted by the internal
galeal muscle, $\tau _{3}$ reflects the decreasing rate of the cross
section of internal galeal muscles along proboscis. Here we
\textrm{choose} $\tau _{3}=50$ mm.

Moreover, the initial conditions $\psi _{0}=\psi (0)$,
$\dot{\psi}_{0}=\dot{\psi}(0)$ and $\ddot{\psi}_{0}=\ddot{\psi}(0)$
in Eq.~\ref{eq} should be given. We note that the internal stipes
muscle and the basal galeal muscle \cite{Krenn2010} (or the proximal
basal muscle and the distal basal muscle \cite{Krenn2000}) locate at
the bottom of the proboscis. It is reasonable to think that they can
adjust the initial conditions $\psi _{0}$, $\dot{\psi}_{0}$ and
$\ddot{\psi}_{0}$ between the proboscis uncoiling and coiling
processes.

\section{Results}

 Now, we will explain how a butterfly manipulates its proboscis.
 According to the experimental observations, the
coiling and uncoiling processes can be divided into four different
movements and we will begin with the loosely coiled state coiled
about 1$\sim$2 loops \cite{Krenn2000}.

\begin{figure}\centering
\includegraphics[width=0.48\textwidth]{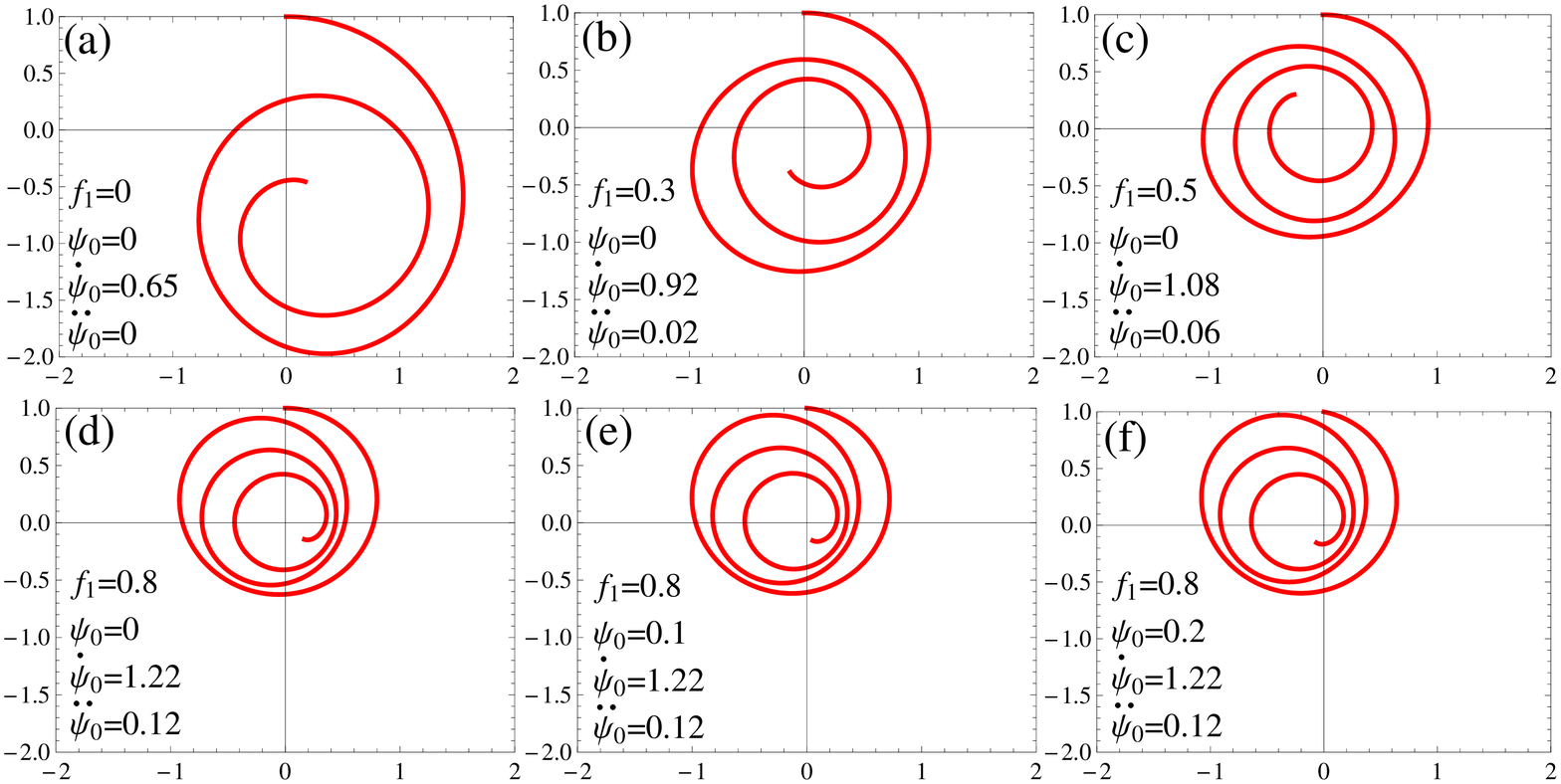}
\caption{\label{fig3} Coiling process from loosely coiled state (a)
to resting position (f), the unit of coordinates axes for each
figure is mm. From (a) to (d) is the coiling process own to the
contraction of the intrinsic galeal muscle and the internal stipes
muscle. The former one increases the tension $f$ and the later
increases the initial conditions $\dot{\psi}_0$ and $\ddot{\psi}_0$.
From (d) to (f), the proboscis is flexed to the resting position due
to the contraction of the internal stipes muscle which induces the
increase of the initial condition $\psi_0$. This two processes can
be coupled together. }
\end{figure}

First movement: coiling from the loosely coiled state to the resting
position. This movement is only driven by muscles and is not related
to the pressure $\Delta P$ \cite{Krenn2000}. Two kinds of muscle are
involved in this movement: the internal galeal muscle and the
internal stipes muscle \cite{Krenn2010}. In our study, the former
one adjusts the line tension $f$ and the later one controls the
initial conditions. In the loosely coiled state, there are $\Delta
P=0$, $f_{1}=0$ and the shape is determined by spontaneous curvature
$C$. From Fig.~\ref{fig3}(a) to Fig.~\ref{fig3}(d), we give a list
of shapes obtained from Eq.~\ref{eq} with the increase of the
tension $f$. In Fig.~\ref{fig3}(a) it is the loosely coiled state
with $\Delta P=0$, $f_{1}=0$ and coiled about 2 turns, which is in
good agreement with the experimental data \cite{Krenn2000}. We find
that the initial conditions should be adjusted simultaneously with
the increase of $f_{1}$, otherwise the shape will coiled irregularly
and self-intersected badly. So it needs the two kinds of muscles
involved in this movement to coordinate with each other. This mutual
coordination can be achieved due to there are plentiful sensilla on
proboscis \cite{Krenn1998}. From Fig.~\ref{fig3}(d) to
Fig.~\ref{fig3}(f), it is only the initial condition $\psi _{0}$ is
changed, which makes the tightly coiled shape is bend to the resting
position. After that proboscis touches on the labium and is fixed by
microtrichia after a slight countermovement which is due to the
muscles have been loosed \cite{Krenn1990,Krenn2010}. The final shape
in Fig.~\ref{fig3}(f) coiled 3$\sim$4 turns is in accordance with
experimental result \cite{Krenn2000}. Actually, the above two
separate processes can be coupled together.

Second movement: uncoiling from the resting position to the loosely
coiled state. When it needs to use proboscis, butterfly unlocks its
proboscis at the tightly coiled resting position and consequently
proboscis uncoils to the loosely coiled state only due to its
elasticity \cite{Krenn1990,Krenn2010}. This process can be taken as
the inverse process from Fig.~\ref{fig3}(a) to Fig.~\ref{fig3}(f).
\begin{figure}\centering
\includegraphics[width=0.48\textwidth]{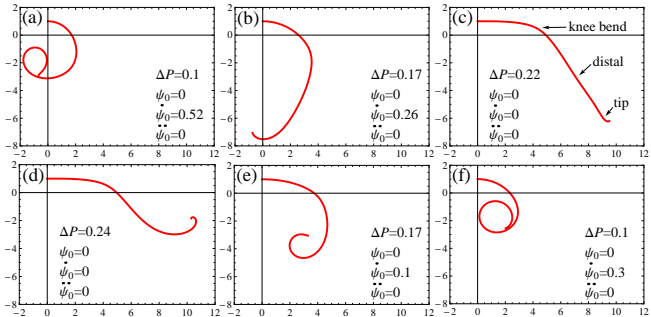}
\caption{\label{fig4} From (a) to (e) is the uncoiling process from
loosely coiled state to extended state, which owns to the increase
of $\Delta P$ and the diereses of initial condition $\dot{\psi}_0$.
From (e) to (f), the proboscis turns back to the resting position
due to the elasticity.  The unit of coordinates axes for each figure
is mm.}
\end{figure}

Third movement: uncoiling from the loosely coiled state to extended
state with a knee bend. This movement is driven by the pressure
$\Delta P$ and the basel galeal muscle \cite{Krenn2000,Krenn2010}.
Choosing loosely coiled state in Fig.~\ref{fig4}(a) as the initial
shape, gradually increasing $\Delta P$ and $\dot{\psi}_{0}$ we
obtain the shapes from Fig.~\ref{fig4}(a) to Fig.~\ref{fig4}(d).
Where we need the initial condition $\dot{\psi}_{0}$ controlled by
the contraction of basel galeal muscle and the pressure $\Delta P$
are increased compatibly. We can see coiled proboscis is opened
gradually in this process. In Fig.~\ref{fig4}(c) and
Fig.~\ref{fig4}(d) there is an evident local bend for each shape at
the position $s=s_{0}$, which just is the knee bend and is due to
the local maximal spontaneous curvature. In Fig.~\ref{fig4}(c) the
shape is composed by two almost strait parts, and between them is
the knee bend. In this case, if butterfly adjusts $\Delta P$, the
tip region is ease to bend backward like Fig.~\ref{fig4}(b) or
forward like Fig.~\ref{fig4}(d). This is in accordance the
experimental phenomenon \cite{Krenn2005} that the tip region is
agile. But if butterfly continually increase $\Delta P$,
Fig.~\ref{fig4}(d) will have more local bends like a wave and it is
very difficult to become to be a totally straight shape with no
local bends. Which implies that the nectar-feeding butterfly with
knee bend on proboscis is difficult to achieve piercing movement
because it needs the proboscis to extend to almost a straight line
\cite{Krenn2008,Krenn2010}. In the later text we will see that if
there is no knee bend, proboscis can easy be unwound totally. The
above whole process is consistent with experimental results in
Fig.~3 of ref.~\cite{Krenn1990}.

Fourth movement: coiling from the extended state to loosely coiled
state. This process ordinarily is only due to the elasticity of
proboscis. But the $\Delta P$ and initial conditions turn back to be
the corresponding values in Fig.~\ref{fig4}(a) that needs a short
time \cite{Krenn1990}. From Fig.~\ref {fig4}(d) to
Fig.~\ref{fig4}(f), we show this coiling process through decreasing
$\Delta P$ and $\dot{\psi}_{0}$. After Fig.~\ref{fig4}(f), the shape
turns back to loosely coiled state in Fig.~\ref{fig4}(a). Experiment
results indicate that the tip region will form a small ring at first
and consequentially the distal region begin to coil
\cite{Krenn1990}. We can see that Fig.~\ref{fig4}(e) and
Fig.~\ref{fig4}(f) are in good agreement with this phenomenon. It
should be noted that this coiling process cannot be simply taken as
the inverse process from Fig.~\ref{fig4}(a) to Fig.~\ref{fig4}(d).
The shapes Fig.~\ref{fig4}(b) and Fig.~\ref{fig4}(e) have the same
$\Delta P$, but the different initial condition $\dot{\psi}_{0}$
leads they have different morphology. Similar results also be shown
in Fig.~\ref{fig4}(a) and Fig.~\ref{fig4}(f). This small difference
is due to the delicate relationship between $\Delta P$ and
$\dot{\psi}_{0}$.

\begin{figure}\centering
\includegraphics[width=0.48\textwidth]{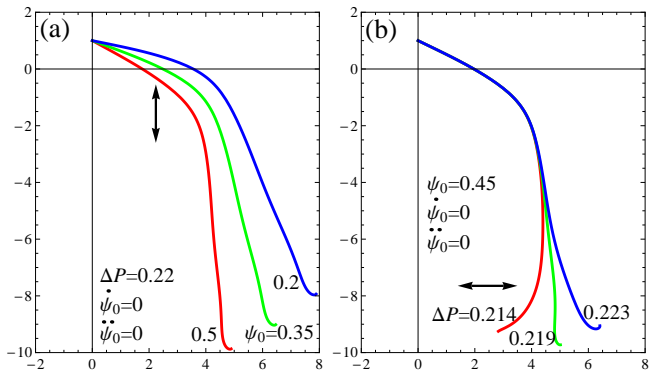}
\caption{\label{fig5} Two particular movements for extended
proboscis with a knee bend, the unit of coordinates axes for each
figure is mm. (a) The up-and-down movement of total proboscis
achieved by adjusting the initial condition $\psi_0$. (b) The
front-and-back movement of distal region induced by the change of
$\Delta P$.}
\end{figure}

Besides the above four movements there are two other movements which
cannot be ignored. It is found that the extended proboscis has two
particular movements: up-and-down movement of the total proboscis
and front-and-back movement of the distal region
\cite{Krenn2008,Krenn2010}. Calculation indicates the former
movement is due to that the initial condition $\psi_0$ is adjusted
by the muscles at the bottom of proboscis. The later movement is
driven by the change of $\Delta P$. Fig.~\ref{fig5} shows this two
movements which can explain the experimental phenomenon well. In
Fig.~\ref{fig5}(a), the change of $\psi_0$ induces total proboscis
moves up and down. In Fig.~\ref{fig5}(b), the change of $\Delta P$
leads to the distal region turning around the knee bend, which can
be taken as the front-and-back movement. The above two movements
make the tip of proboscis can reach a much bigger area region so
that butterfly has the possibility to suck more nectar. However,
only changing the line tension also can induce the similar
front-and-back movement in Fig.~\ref{fig5}(b). (It needs $f_1$ to be
negative in Eq.~\ref{f} in this case.). Actually, experimental
investigation indicates the front-and-back movement is caused either
by pressure or galeal muscle \cite{Krenn1990}. It also finds that
there are slight difference between the two kinds of galeal muscles
in the knee bend region \cite{Krenn1990}. These phenomena imply the
line tension in Eq.~\ref{f} which is induced by the galeal muscles
locating along proboscis is possible to have complicated
distribution feature like the spontaneous curvature in
Eq.~\ref{SC1}. We also can presume the butterfly can adjust the
local line tension by controlling the extension and constriction of
local galeal muscles along proboscis to achieve a more complicate
movement. In the Movie S1, we give a full circle of the coiling and
uncoiling processes containing the front-and-back and up-and-down
movements.

\begin{figure}\centering
\includegraphics[width=0.48\textwidth]{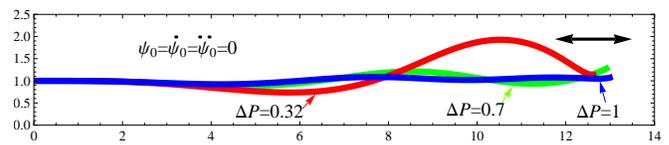}
\caption{\label{fig6} Piercing action induced by increase of $\Delta
P$ for proboscis without knee bend. The unit for coordinates axes is
mm.}
\end{figure}

There are other kinds of butterflies that their proboscis have no
evident knee bend but they have a particular ability: piercing. We
note that the piercing action occurs if and only if proboscis
extends to an almost straight line \cite {Krenn2008,Krenn2010}.
Considering our former result that proboscis which has knee bend
cannot be completely extended, we can believe that kneed bend is
disadvantageous to the piercing action. Thus, the proboscis of
fruit-piercing butterfly has no evident knee bend. In order to
demonstrate how the piercing action occurs, we choose the following
spontaneous curvature which has no local maximum to induce the knee
bend
\bea
\label{piercing}      C=0.1\exp[-s/50]+0.4.
\eea
There isn't any configuration parameters of proboscis of
fruit-piercing butterfly, we simply choose the results in
Fig.~\ref{fig2}. Calculation shows piercing action can be archived
easily in this case. Fig.~\ref{fig6} shows the piercing action
induced by the increase of internal pressure. We can see an evident
straight movement of the tip along the direction of the proboscis,
which helps proboscis to pierce into fruit and other softer targets.
The final state with $\Delta P=1$ in Fig.~\ref{fig6} is nearly a
straight line. Actually choosing the spontaneous curvature in
Eq.~\ref{SC1} also can achieve the straight shape. But we find it
needs a much higher pressure: $\Delta P>10$. Compered with the
straight shape in Fig.~\ref{fig6} only needs $\Delta P=1$, so higher
pressure is not convenient (or impossible) for butterfly. So, knee
bend is no need for a fruit-piercing butterfly.

\section*{Discussion}

Through our analysis, we find the achievement of whole coiling and
uncoiling movent subtly depends on these parameters: $\kappa $, $C$,
$\sigma $ and $f$. In Fig.~\ref{fig2}, we used the quintic functions
of $s$ to fit $\sigma $ and $\kappa =bh^{3}$. We also tried other
fitting functions with lower orders, but we found quintic functions
can make sure the coiled proboscis is more regular than others. The
$\sigma $ and $\kappa =bh^{3}$ in Fig.~\ref{fig1} seem to reach
their local mutation points at the same length $s\simeq 4$ mm which
is close to the knee bend location $s=4.44$ mm. But our calculation
indicates the local mutation of $\sigma$ and $\kappa =bh^{3}$ can
not induce knee bend if there isn't local maximum of $C$. Even if
choosing invariable $\sigma$ and $\kappa$, we also can not attain
knee bend if there isn't local maximum of $C$. So we make sure it is
the local maximum of spontaneous curvature that leads the knee bend.
Further, our study reveals that $\Delta P$ and the initial
conditions should be changed compatibly, which needs the different
kinds of muscles involved in each movement to coordinate with each
other. Besides the plentiful sensilla on proboscis which can help
butterfly to achieve this mutual coordination, there are other
particular movements that make for this cooperation. Experimental
details indicate that the uncoiling movement is a step-by-step
process resulting from the step-by-step increase of the hemolymph
pressure \cite{Krenn1990}. So the change of $\Delta P$ is
discontinues, which implies there is a short stagnant time after
each change of $\Delta P$. In this short buffer time, the initial
conditions can be adjusted by galeal muscles to the optimal values
to cooperate with the pressure and consequently whole shape can
reach the equilibrium state.

It is believable that the construction and behavior of proboscis are
due to the evolution. So, the proboscis of nectar-feeding butterfly
has knee bend that is in order to adapt the environment.
Fig.~\ref{fig4} has indicated that knee bend will induce proboscis
has two particular movements: up-and-down movement and
front-and-back movement, and the couple of them can ensure the tip
of proboscis to reach a much bigger area than a strait proboscis
without knee bend can. Apparently, the more reachable area the more
possibilities to get nectar. However, this advantage will be a
disadvantage when proboscis is used to pierce into a target, because
it needs a much higher pressure when compared with a no knee bend
proboscis. In order to avoid this disadvantage, fruit-piercing
butterflies choose the proboscis without knee bend to obtain the
piercing ability. But why the knee bend locates at the $35\%$ length
from the bottom of proboscis is an interesting problem. May be this
structure can well adapt the shape of flowers or it is the result of
whole body structure optimization. For instance, in Fig.~\ref{fig1}
the knee bend locates at $30\%$ length from the bottom of proboscis,
which deviates the value $35\%$ in Ref.~\cite{Krenn2000}. So we
think this proportion is variational for different kinds of
butterfly.

There are other animals that have long proboscises \cite{Photo},
such as the marine snail \cite{Alan,Greene,Salisbury} and elephant
\cite{Jeheskel}. The proboscis of marine snail is used to catch fish
and the details of its movements still are unknown. As to the
elephant, proboscis is the most important and versatile appendage
and has higher flexibility to achieve complex movements relying on
it has more than 15000 muscle fascicles. In our study, we choose a
simpler line tension in Eq.~\ref{f} to reflect that the muscles
along proboscis can adjust the shape of proboscis. But we believe
the line tension can be adjusted locally in some cases by butterfly
and other animals to achieve complex movements. The more muscles it
has, the more complex movements it can achieve. So, to ascertain the
line tension $f$ is an important step to investigate the complex
behaviors of the proboscis with plentiful muscles. But this is very
difficult, because the feedback control loop composed by the brain,
nerves and abundant sensilla on proboscis which manipulates the line
tension is unquantifiable at present time. Whereas, our study
provide a revelatory method to investigate the mechanical behaviors
of proboscis and other Q1D biologic structures, such as the growth
of twigs of plants and hairs. Twigs of plants often have variable
cross section, so they probably have nonlinear $\kappa $, $\sigma $
and $C$. Although they have no muscles, the increase of internal
hydraulic pressure can induce the curved twigs to straight and
upwardly grow up.

In summary, we revealed that, in the coiling and uncoiling process
of butterfly proboscis, the function of muscles located the bottom
of proboscis is to adjust their initial states that govern the
corresponding movement of the coiling and uncoiling processes
respectively, the function of muscles existed along proboscis is to
control the line tension, and the knee bend determined by the local
maximal spontaneous curvature is a disadvantage for piercing action.
The knee bend induces that the tip of proboscis has two particular
movements: the up-and-down movement and the front-and-back movement,
and the combine of them leads to the tip can reach a bigger area
region when compared with the no knee bend state. Our study
indicates that the nonlinear construction parameters: $\kappa $,
$\sigma $ and $C$ of proboscis make it is very difficult to
accurately control whole coiling and uncoiling process. But we
believe these parameters have been optimized by long time of
evolution to insure animals can obtain the maximal advantage. Two
typical advantages present that proboscis has higher retractility
and particular movements, which can help butterfly to get more food.
Beside Lepidoptera, our model gives a general method to study the
mechanical behaviors of long proboscis of other animals as well as
twigs of plants. Moreover, our method and results can be used to
design bionic devices.

ACKNOWLEDGMENTS. This work is supported by the National Natural
Science Foundation of China Grants 11304383 and 11374237.

\end{document}